\documentclass[lettersize,journal]{IEEEtran}
\usepackage{amsmath,amsfonts}
\usepackage{algorithmic}
\usepackage{algorithm}
\usepackage{array}
\usepackage[caption=false,font=normalsize,labelfont=sf,textfont=sf]{subfig}
\usepackage{textcomp}
\usepackage{stfloats}
\usepackage{url}
\usepackage{booktabs}
\usepackage{verbatim}
\usepackage{graphicx,lscape}
\usepackage{multicol, blindtext}
\usepackage{blindtext}
\usepackage{cite}
\usepackage{tabularx}
\usepackage{float}
\usepackage{stfloats}
\usepackage{lipsum}
\usepackage{color}

\begin{document}

\title{{Simultaneous Transmitting and Reflecting-Reconfigurable Intelligent Surface in 6G: Design Guidelines and Future Perspectives}}

\author{Waqas Khalid, Zeeshan Kaleem, Rehmat Ullah, Trinh Van Chien, Song Noh and Heejung Yu
 \vspace{-0.8cm}       
\thanks{Waqas Khalid and Heejung Yu (corresponding author) are with Korea University, Sejong, South Korea; Zeeshan Kaleem is with COMSATS University Islamabad, Wah Campus, Pakistan; Rehmat Ullah is with Cardiff Metropolitan University, Cardiff, UK; Trinh Van Chien is with  Hanoi University of Science and Technology, Vietnam; Song Noh (corresponding author) is with Incheon National University, Incheon, South Korea.}
}



\maketitle

\begin{abstract}
Reconfigurable intelligent surfaces (RISs) have been considered as a promising technology for the sixth-generation (6G) wireless networks that can control wireless channels in a desirable way and significantly enhance the network performance. Simultaneous transmitting and reflecting-RISs (STAR-RISs) can overcome limitation of reflecting-only RISs by leveraging the higher design flexibility and full-space coverage. Despite the benefits, the modeling and analysis of STAR-RISs are complicated because of various control parameters for both transmission and reflection links. In this article, a general framework to facilitate the design and implementation of STAR-RISs in 6G scenarios and network topologies is presented. We provide a systematic introduction to the STAR-RIS operating protocols for different communication modes and discuss recent efforts to identify the research progress and combination solutions. Finally, we provide the design concepts, research challenges, potential solutions, and future directions related to the channel modeling, channel estimation, hardware implementations, modeling and limitations, and optimization.
\end{abstract}

\section{Introduction}

{The conventional transmission strategy focuses on the design of transceivers} by leveraging multiple antennas, effective encoding and decoding procedures, and advanced communication protocols. In future sixth generation (6G) networks, reconfigurable intelligent surfaces (RISs) are envisioned to control wireless channels {and provide solutions} to radio-frequency (RF) impairment {and signal propagation. An RIS ensures} wireless connectivity via non-line-of-sight (nLoS) links under harsh propagation conditions, e.g., {LoS link blockage} or shadowing, and extends the communication distance via controlled scattering and multipath components. {An RIS enables solutions for better coverage and transmission reliability without sophisticated signal processing {or RF operations}. However, signal manipulation is widely investigated for reflecting-only RISs \cite{ref1}. The topological constraint, i.e., the same side deployment of the transmitter and receiver, for reflecting-only RISs causes performance degradation. A reflecting-only RIS reflects the signals within front half-space and blocks the signals to the users behind the RIS, which results in degraded performance for blocked users. The location constraint for the nodes limits the service coverage to $180^o$, restricts the effectiveness of the RIS, and hinders 6G small-cell implementations. To overcome such limitations, simultaneous transmitting and reflecting-RISs (STAR-RISs) has been introduced \cite{ref2}.}

{A STAR-RIS offers flexible implementation to provide both transmission (T) and reflection (R) signals. The STAR-RIS establishes a full-space (i.e., $360^o$) wireless coverage via serving users located on both sides. The location of a user determines that its signal is reflected or transmitted.  Thus, the STAR-RIS provides a higher design flexibility with more degrees of freedom for signal manipulation. The STAR-RIS introduces more adjustable parameters and both T and R communication links. Therefore, the previous results of conventional RISs do not directly apply to STAR-RISs. The modeling and analysis of STAR-RISs are still in an early stage. Motivated by the above discussion, we present a general framework to facilitate the design and implementation of STAR-RISs in future 6G deployments and discuss the effective STAR-RIS designs under variable network topologies and application scenarios.} {The contributions of this article are summarized as follows:
\begin{itemize}
\item A systematic overview on the STAR-RISs is presented by discussing the differences with other variants of RISs. The benefits and drawbacks of the operating protocols of STAR-RIS are examined and their performance is evaluated. We also present a research review to discuss the recent efforts and identify the research gaps. Furthermore, the promising solutions for STAR-RISs to achieve diverse communication objectives in 6G scenarios are discussed.
\item Key ideas, design issues and potential solutions of STAR-RISs are investigated. Based on the investigation, valuable insights and future directions are provided with respect to the channel modeling, channel estimation, beamforming and element reconfiguration, hardware implementations, modeling and limitations, and optimization.
\end{itemize}
}

{The article is organized as follows. Section \ref{section2} presents a systematic overview and research review. Section \ref{section3} discusses the research challenges, potential solutions, and future directions from the perspective of the design and implementation of 6G STAR-RISs. Section \ref{section4} concludes the article.}

\begin{figure}[t!]
\centering
\includegraphics[width=3in,height=1.8in]{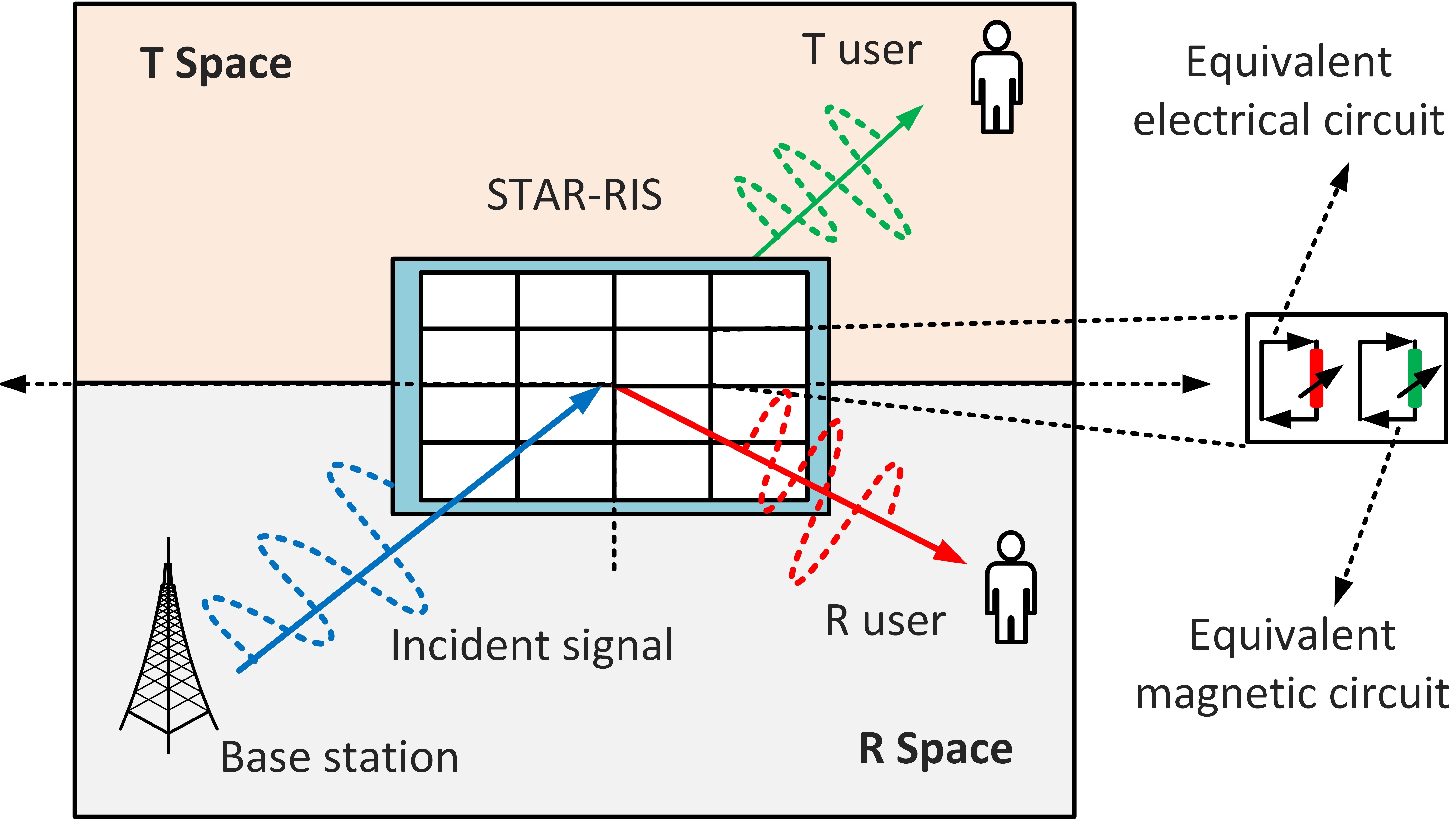}
\caption{Concept of full-space service coverage by STAR-RIS.}
\label{fig1}
\end{figure}

\section{STAR-RIS: A Systematic Overview and Research Review}

\label{section2}

\begin{figure*}[t!]
\centering
\includegraphics[width=4in,height=3.4in]{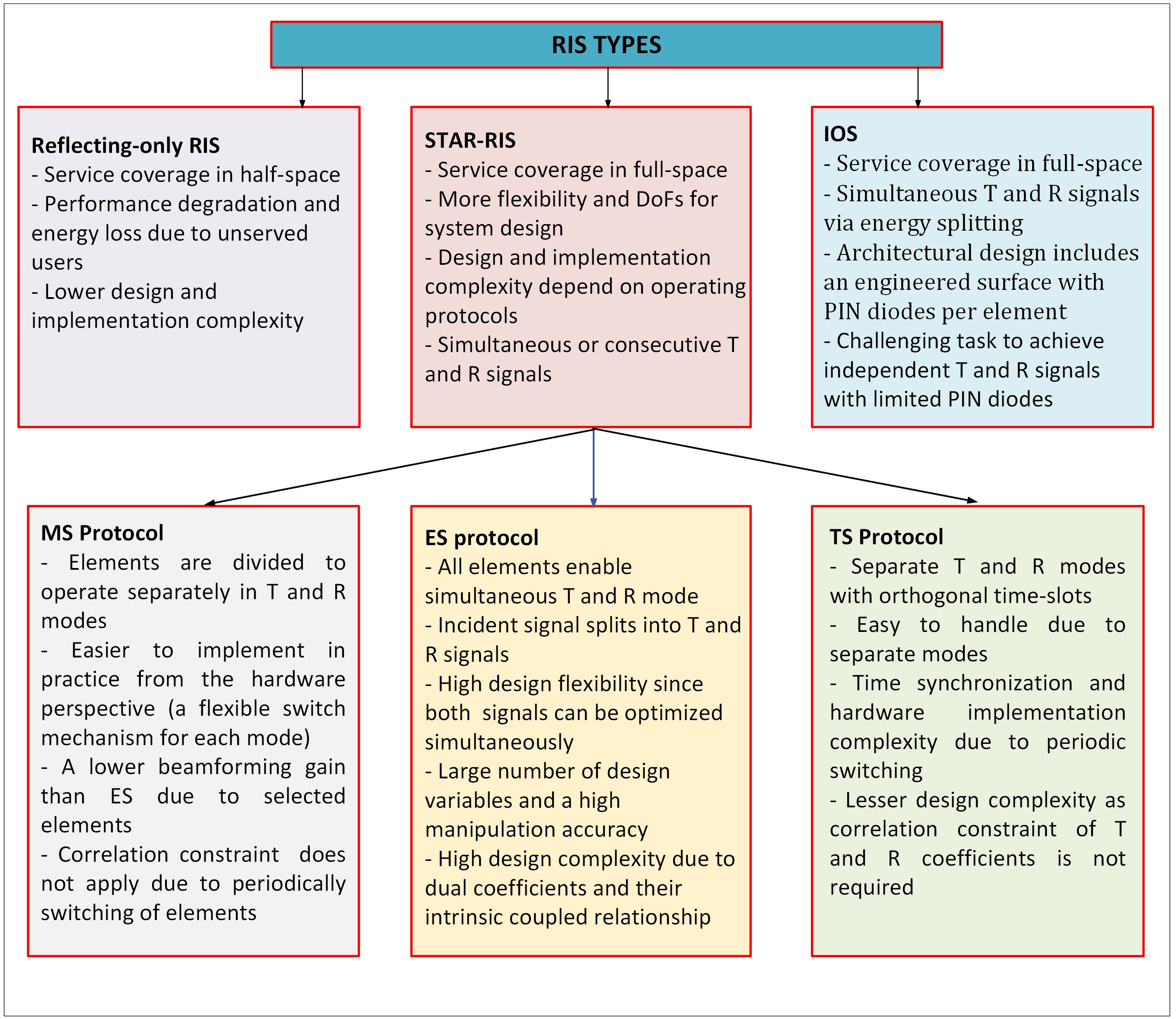}
\includegraphics[width=3.0in,height=3.4in]{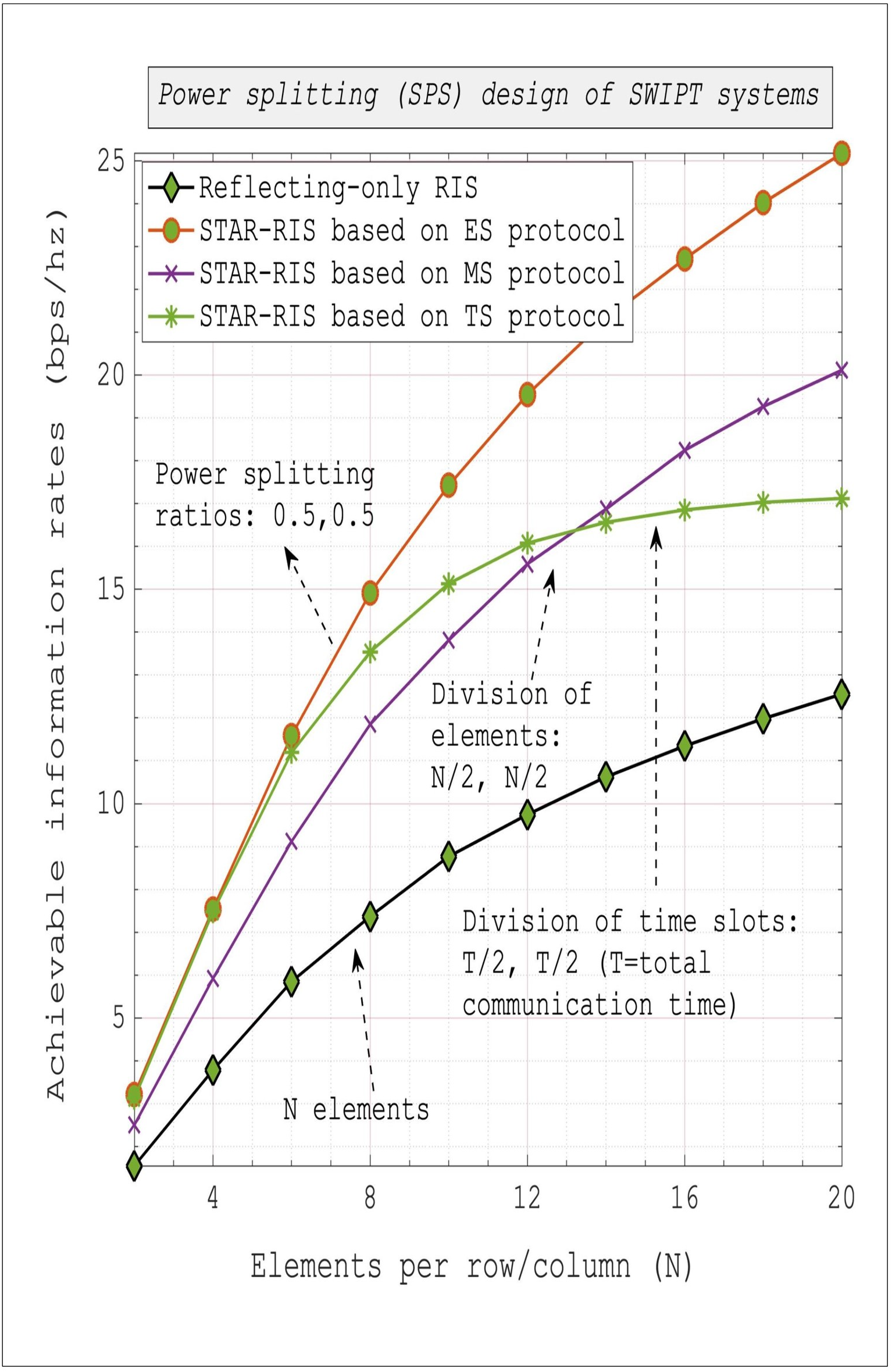}\\
\hspace{0.5in}(a) \hspace{3.5in}(b)
\caption{{Comparison between the STAR-RIS operating protocols and other RIS types: (a) Details and (b) Achievable rate performance.}}
\label{fig1a}
\end{figure*}

{A STAR-RIS is composed of many subwavelength  reconfigurable elements. As compared to other variants, i.e., holographic MIMO surfaces and RISs, the STAR-RIS facilitates simultaneous or consecutive T and R signals with coupled or independent controls. Both STAR-RIS and intelligent omni-surface (IOS) are innovative metasurface implementations to serve users in full-space coverage. However, IOS and STAR-RIS are different in terms of the architectural design and communication protocols. An IOS splits the incident signal into T and R signals with different amplitudes and phases, and provides the simultaneous T and R mode. A STAR-RIS supports both electric and magnetic currents to produce T and R signals, as shown in Fig. \ref{fig1}.} The three operating protocols for STAR-RISs, i.e., energy splitting (ES) \cite{ref2}, mode selection (MS) \cite{ref3}, and time splitting (TS) \cite{ref4}, {unlock different communication modes, including T, R, and simultaneous T and R modes}. The important aspects of the operating protocols are shown in Fig. \ref{fig1a}(a), and are highlighted as follows \cite{ref2,ref3,ref4,ref5}:
\begin{itemize}
\item ES protocol: All elements enable simultaneous T and R mode. The amplitude and phase variations can be jointly optimized in element-wise T and R coefficients.
\item MS protocol: The elements are partitioned to operate separately in T and  R modes. Thus, the number of selected elements and the corresponding phase shift variations can be jointly optimized.
\item TS protocol: The elements are periodically switched in the time domain for T and R modes. Thus, the orthogonal time slots for both modes can be optimized.
\end{itemize}


 \begin{table*}[t!]
{\caption{Recent efforts toward the realization of STAR-RISs.}
\centering
\begin{tabular}{|p{.4cm} | p{1.4cm}|   p{1.5cm}  p{1.9cm}  p{1.5cm} | p{2.6cm}| p{.9cm} | p{.9cm}  p{1.3cm}  p{1.7cm}| }
\toprule

\textbf{Ref.} & 

\textbf{System model }

 --------------

Operating protocol (network) &

 \;\;\;\;\;\;

  \;\;\;\;\;\;
  
----------------
  
 Performance metric

 &\textbf{Design and optimization}
 
-------------------

 Design variables &

 \;\;\;\;\;\;
 
 \;\;\;\;\;\;
 
 --------------
  
   Optimization methodology  &\textbf{Insight remarks} &\textbf{CSI estimation } &

  \;\;\;\;\;\;

  \;\;\;\;\;\;
  
  \;\;\;\;\;\;

--------- 
   
   Channel model & \textbf{Modeling and limitation}
  
  -------------- 
   
   Hardware model & \;\;\;\;\;\;

  \;\;\;\;\;\;
  
  \;\;\;\;\;\;

---------------
   
 Hardware limitation \\   \midrule

  \; \cite{ref3} & 
 
ES, TS, MS (MIMO)
 
 &    Sum rates  & Precoding matrices and STAR-RIS coefficients & BCD (analytical)& Design the beamforming matrices in a sub-optimal manner&\;\;\; \textbf{$\times$} & Rayleigh fading &  Phase shift & Ideal architecture
\\
  \midrule
 \; \cite{ref4}
 & 
 
 ES, MS (MIMO)

 & Coverage probability & STAR-RIS coefficients &\;\;\;\;\;\;

 \;\;\;\;\;\;----&  Incorporate the statistical CSI for T and R users &\;\;\; \textbf{$\times$} & Correlated Rayleigh fading &  Phase shift& Discrete amplitude and phase shifts
 \\
 \midrule
 \; \cite{ref5}
  & 
 
 ES, TS (SISO)

 &Channel estimation error & Training pattern matrices & Sub-optimal methods (analytical)
 &  TS outperforms ES in terms of uplink channel estimation &\;\;\; \textbf{$\checkmark$}& Rayleigh fading &  Phase shift &  Coupled phase shifts\\
   \midrule
 \; \cite{ref8}
 & 
 
ES (SISO-NOMA-CoMP)&  Achievable

\;\;\;\; rates & STAR-RIS coefficients &\;\;\;\;\;\;

 \;\;\;\;\;\;----& Eliminate the inter-cell interference and boost the desired signal &\;\;\; \textbf{$\times$} & Rayleigh fading& Phase shift &Ideal architecture\\

 \midrule
 \; \cite{ref9}
 & ES (MISO-NOMA) & \;\;\; EE  &  Precoding matrices and STAR-RIS coefficients & DDPG (ML)& Solve the fractional objective function with non-convex constraints&\;\;\; \textbf{$\times$} & Rayleigh fading &  Phase shift& Ideal architecture
 \\
     \midrule
 \; \cite{ref10}
 &  ES (SISO-NOMA) &  Outage probability &  Transmit SNR and a target SNR at receiver & \;\;\;\;\;\;

 \;\;\;\;\;\;----
 & Incorporate the coupled phase shifts of T and R users &\;\;\; \textbf{$\times$}& Rician fading &  Load impedance & Coupled phase shifts
  \\
\bottomrule
\end{tabular}
\label{table2}}

\end{table*}


{Achievable rates in Fig. \ref{fig1a}(b) are extended from the previous work \cite{ref6} for the ES, MS and TS operating protocols and provide a comparison with a reflecting-only RIS. We consider correlated Rayleigh fading channels, multicast transmission (i.e., the same information is sent to for T and R users), and power splitting in simultaneous wireless information and power transfer (SWIPT) design. The detail parameters used for simulations are given in \cite{ref6}. The numerical results show that the optimal selection and superiority of the operating protocols depend on the communication objective. By altering the parameters of a STAR-RIS (e.g., an energy splitting ratio in ES protocol, the number of elements in MS protocol, and the duration of orthogonal time slots in TS protocol), the STAR- RIS can satisfy the demands of both T and R users and can also demonstrate a simple RIS scenario (to serve either T or R user). The selection strategy determines the hardware and design complexity of STAR-RISs as well as performance gain.}

{Through intelligent placement and effective beamforming designs in a full space smart radio environment offered by the STAR-RISs, 6G wireless paradigm can deliver its full potential in terms of signal boosting for millimeter-wave (mmWave) and terahertz (THz) signals, power consumption reduction, throughput enhancement, interference suppression, secure transmission (via physical layer security), {non-orthogonal multiple access (NOMA)}, and SWIPT. Inspired by the aforementioned benefits, STAR-RISs can enable novel application scenarios, e.g., STAR-RISs can be deployed on the existing structures (such as facades, walls and ceilings of the buildings) to provide cascaded indirect links to both indoor and outdoor users}. However, research efforts from industry and academia have mainly focused on the design and optimization of the conventional RISs. {In this context, unmanned aerial vehicle (UAV)-RIS can alter the wireless propagation through maneuver control and signal manipulation in full-space. However, UAV-RIS poses crucial challenges, including complex channel estimation and modeling, and more vulnerability to the security attacks \cite{ref7}.} {There have been initial studies that covered STAR-RIS-aided systems and discussed their potential benefits and combination solutions \cite{ref3,ref4,ref5,ref8,ref9,ref10,ref11}.} Table \ref{table2} summarizes the relevant research contributions.

{STAR-RISs have been studied in next-generation wireless systems \cite{ref3,ref4,ref5}. In \cite{ref3}, a conventional (sub-optimal) block coordinate descent (BCD) algorithm was employed for solving the sum rate optimization problem in STAR-RIS-aided MIMO networks for unicast and broadcast communications. In \cite{ref4}, the authors derived a coverage probability expression for a STAR-RIS-aided massive MIMO system under correlated fading and phase-shift errors. Moreover, \cite{ref5} presented the uplink channel state information (CSI) acquisition techniques for the TS and ES protocols.} {Several studies have shown that STAR-RIS provides a natural fit for NOMA to meet stringent future 6G requirements \cite{ref8,ref9,ref10}. In particular, NOMA improves spectral efficiency and connectivity, and STAR-RIS extends the full-space coverage and facilitates the basic NOMA principle. In particular, the channel gains can be adjusted via ES and MS protocols to provide flexible decoding orders according to the requirements of T and R users.} In \cite{ref8}, the transmission design of a STAR-RIS-NOMA-coordinated multi-point (CoMP) system was addressed to simultaneously eliminate inter-cell interference and boost the desired signal. In \cite{ref9}, the energy-efficiency (EE) maximization problem for a STAR-RIS-NOMA network was solved using machine learning (ML). Specifically, a deep reinforcement learning approach, i.e., a deep deterministic policy gradient (DDPG), was considered. Moreover, correlated T and R phase shifts were proposed for a STAR-RIS-NOMA system and the outage probability and power scaling law were derived \cite{ref10}. 

Despite the aforementioned research efforts, the study and evaluation of 6G STAR-RISs are in an early stage. {The design and implementation of STAR-RISs under variable network topologies and application scenarios require further investigation.} Specifically, the synergy between different communication modes and the inherent complexity with more adjustable parameters have not been studied extensively.

\section{Research Challenges, Potential Solutions, and Future Directions} 

\label{section3}

\begin{table}[t!]
{\caption{ Details of channel models and approximations for STAR-RISs. }
\centering
\begin{tabular}{ |p{1.3cm}| p{1.6cm}| p{4.8cm}| }
\toprule
\textbf{Methods}	& \textbf{{Physics-based models}} & \textbf{Details} \\
\midrule

\;\;\;\;\;\;\;\;\;

\;\;\;\;\;\;\;\;\;

\;\;\;\;\;\;\;\;\;

\;\;\;\;\;\;\;\;\;

\;\;\;\;\;\;\;\;\;

\;\;\;\;\;\;\;\;\;

\;\;\;\;\;\;\;\;\;

Exact &

\;\;\;\;\;\;\;\;\;

\;\;\;\;\;\;\;\;\;

\;\;\;\;\;\;\;\;\;

\;\;\;\;\;\;\;\;\;

Ray-tracing-based

\;\;\;\;\;\;\;\;\;

\;\;\;\;\;\;\;\;\;

\;\;\;\;\;\;\;\;\;

\;\;\;\;\;\;\;\;\;

---------------

\;\;\;\;\;\;\;\;\;

Huygens– Fresnel–based &
--  Compatible {with the} phase shift and load impedance models

--  Consider the principles of geometrical optics for ray-element interaction

-- Incorporate the physical factors, such as environment geometry, frequency, and locations of nodes

-- Applicable to the far-field region under certain conditions

-- Simple signal power calculation

\;\;\;\;-------------------------------------------

--  Applicable to the near-field region

--  Analyze the wavefront transformation function of the STAR-RIS

--  Limited to LoS and free-space scenarios

--   {Challenge to select a} proper wave-front to characterize T and R coefficients  \\
\midrule
 
 \;\;\;\;\;\;\;\;\;

\;\;\;\;\;\;\;\;\;

\;\;\;\;\;\;\;\;\;

\;\;\;\;\;\;\;\;\;

\;\;\;\;\;\;\;\;\;

\;\;\;\;\;\;\;\;\;

\;\;\;\;\;\;\;\;\;

Approximate&

\;\;\;\;\;\;\;\;\;

 \;\;\;\;\;\;\;\;\;

Central limit

 \;\;\;\;\;\;\;\;\;

  \;\;\;\;\;\;\;\;\;
  
---------------

\;\;\;\;\;\;\;\;\;

M-fold convolution

---------------

\;\;\;\;\;\;\;\;\;

 \;\;\;\;\;\;\;\;\;

Curve-fitting

&
--  Tractable and accurate expressions for the performance metrics

--  Limited to a large number of elements and a low-medium SNR region

--  Not suitable for the diversity gains

\;\;\;\;-------------------------------------------

--  Diversity analysis to evaluate the performance under high SNR 

--  {Low-complex asymptotic analysis}

\;\;\;\;-------------------------------------------

--   Wider applications (e.g., fewer elements of STAR-RIS and 6G multi-cell scenarios)

--  {Low-complex tractable derivation}

--  Adjustable curve-fitting functions with a challenge of a suitable selection 

--  Not suitable for a high SNR region\\ 
\bottomrule
\end{tabular}
\label{table3}}
\end{table}

\subsection{Channel Modeling for STAR-RISs}

{The channel models for the STAR-RISs integrate the hardware implementations with the communication theory and find out the received signal power for T and R users.} Table \ref{table3}  emphasizes the details of the channel modeling. {The channel models are required to incorporate the necessary propagation statistics, e.g., path loss, shadowing, and small-scale fading \cite{ref12}.  The previously reported accurate but tractable path loss models (for RISs) can be extended for STAR-RISs. The analytical framework can be formulated to compute T and R communications incorporating the impact and scaling laws of design parameters (e.g., the physical size, and the transmission distances). We present the novel physics-based approaches that can be used to develop the channel models for the STAR-RIS from perspective of the far-field and near-field regions (based on the changing behavior of power with distance for the specific configuration). In this context, ray-tracing technique can be adopted for far-field channel models. For near-field users, the wave-front transformation function of STAR-RISs can be studied by} Huygens–Fresnel principle, and Fresnel–Kirchhoff diffraction formula can be extended to calculate the channel gains. For approximation analysis, the central-limit, M-fold convolution, and curve-fitting models can be adopted. {The central-limit model is a mathematical tool to derive the channel model as a summation of random variables. The M-fold convolution method can derive the diversity order (i.e., an asymptotic reliability metric) to evaluate a high SNR performance. Furthermore, different curve-fitting functions can be exploited. In addition, gamma distribution can be considered for the approximation of the received signal.}

\subsubsection{Open Issues and Future Directions}
{In the literature, design and physical properties are generally discussed. However, there is a lack of physical implementations-compliant and mathematical-tractable channel models for 6G STAR-RISs.
\begin{itemize}
\item Numerically reproducible channel models to calculate T and R coefficients and derive the physical-layer performance with the adjustable parameters for the propagation environment and geometric setting (e.g., size and placement plane of the STAR-RIS) are still in their infancy.
\item Channel models are required for different network scales (e.g., single- and multi-user), deployment scenarios (e.g., indoor and small cell), channel conditions (e.g., static and time-varying), and antenna settings (e.g., single antenna and massive MIMO).
\item Outdoor application scenarios are generally formulated. Near-field propagation (e.g., indoor scenarios) and wireless solutions must be discussed extensively considering implementations with a large aperture (in terms of wavelengths) or using higher carrier frequency (e.g., mmWave and THz).
\item Since the exact fading distribution to characterize the system design and performance optimization will be non-trivial in complex environment, the derivation of appropriate mathematical approximations will be required.
\end{itemize}}

\subsection{Channel Estimation Designs for STAR-RISs}

Timely and accurate CSI acquisition is indispensable for a {substantial performance gain} in conventional RIS- and STAR-RIS-aided networks. However, in practical terms, it is a nontrivial task to acquire perfect CSI with negligible training overhead, complexity, and on-board signal processing, even for a conventional RIS. This is because of many reflecting elements with hardware constraints and non-linear characteristics. The passive RIS lacks RF chains to transmit or receive pilot signals to estimate unknown channel coefficients. Most of the literature studies for the conventional RIS assume the CSI availability {(at the transceiver, and RIS)}. The channel estimation methods for both cascaded and separate channels have been investigated under different system and channel models. The multidimensional structure of the received signals using parallel factor (PARAFAC) tensor decomposition can be leveraged to estimate the high dimensional RIS channels with low complexity, and high accuracy and stability in various scenarios \cite{ref13}. The channel features, e.g., high-dimensional but quasi-static (base station-RIS channels), mobile but low-dimensional (RIS-user channels), and low-rank and block sparsity (mmWave channels) can also be exploited. In addition, deep learning (DL) approaches can estimate multiple large RIS channels for the general systems or complicated channel conditions. However, the disadvantages of offline trained networks are the higher hardware complexity, power consumption, and a mismatch with the real-world conditions. The channel estimation methods for RIS cannot be directly applied to STAR-RIS because both T and R patterns need to be designed for STAR-RIS in contrast to designing a single T (or R) pattern for reflecting-only (or transmission-only) RIS. The channel estimation for STAR-RISs revolves around hardware designs and operating protocols. The estimation performance in terms of the training overhead, accuracy and computational complexity is generally influenced by the number of T and R users, elements, and operating protocols. The channel estimation designs for STAR-RISs are still in an early stage and are presented under the top-level taxonomy of model-based (e.g., signal-processing), and data-driven-based (e.g., ML) methods.

\subsubsection{Model-based Methods}

{Least square (LS), linear minimum mean-squared-error (LMMSE), compressed sensing, and matrix factorization, can be adopted to estimate STAR-RIS channels. {Specifically, the closed-form analytical expressions of the pilot sequences and training patterns of the T and R users can be constructed. Different strategies, e.g., grouping of elements, (with a common T or R coefficient), low-rank channel covariance, and use of common channel (e.g., BS-STAR-RIS), can be applied to provide a flexible trade-off between training overhead and beamforming performance.} The simple/low-complex pilot-based channel estimation methods, including LS/LMMSE, are applicable to overdetermined linear signal models, and a wider range of system configurations (e.g., single-/multi-user) and antenna configurations (e.g., SISO/SIMO/MISO/MIMO). However, LS/LMMSE-based estimation suffers from performance degradation under non-linear and non-stationary channels. Compressed sensing exploits the sparse and low-rank mmWave and THz STAR-RIS channels, reducing the training overhead and performance error, and is applicable to under-determined linear signal models. Furthermore, matrix factorization methods decompose the high-dimensional aggregate STAR-RIS channels (e.g., MIMO and multi-user) into a series of lower-dimensional sub-channels with a lower training overhead. The compressed sensing and matrix factorization offer a higher computational complexity than LS/LMMSE. The channel estimation of the STAR-RIS operating protocols can be generally discussed as:}

{\begin{itemize}
\item \textit{ES protocol}: Simultaneous estimation of T and R cascaded channels requires a lesser pilot overhead at the expense of a higher design complexity. The intricately coupled T and R channel coefficients per element cause performance degradation. To solve a challenging CSI acquisition for the ES protocol, a solution based on independent arbitrary T and R channel coefficients can be first obtained which can then be utilized to develop a nearly-optimal solution with a coupled phase-shift model.
\item \textit{MS protocol}: The MS protocol requires lesser pilot overhead and computational complexity because the correlation relationship of the T and R coefficients is not required with the separate estimation of the channels.
\item \textit{TS protocol}: The separate estimation of T and R channels (in different time slots) results in a higher accuracy at the expense of a higher pilot overhead.
\end{itemize}}

\subsubsection{Data-driven-based Methods}

{Data-driven approaches cover different wireless areas, such as channel estimation, channel tracking and prediction, and beamforming. Specifically, the feature extraction and nonlinear mapping capabilities can be applied for channel estimation. ML-based approaches can learn the STAR-RIS channels efficiently for any network configuration and signal model. The conventional approach relies on the channel model, which can be complex and costly to estimate, and can only provide an approximation to reality. On the other hand, ML-based approaches can provide efficient estimation solutions with low training overhead and high accuracy without explicit channel models. ML-based approaches can demonstrate better estimation performance in the real-time under complex conditions, such as statistical correlation and non-stationary channels.}

\subsubsection{Open Issues and Future Directions}
{\begin{itemize}
\item The key aspects of the model-based and data-driven channel estimation approaches for the STAR-RIS are to derive the explicit channel models and determine the accurate but low complex learning algorithms, respectively.
\item The channel estimation methods with the optimal performance-overhead tradeoff will be crucial for 6G STAR-RIS operating protocols.
\item ES protocol can be deployed in scenarios with the low mobility and channel variation to tackle the channel estimation for the intricately coupled T and R signals.
\item The joint design of the training signal and channel estimation algorithm is nontrivial. The optimization of a training signal under a chosen estimation algorithm can be an efficient approach.
\end{itemize}}

  \begin{figure*}[t!]
\centering
\includegraphics[width=6.0in,height=6.0in]{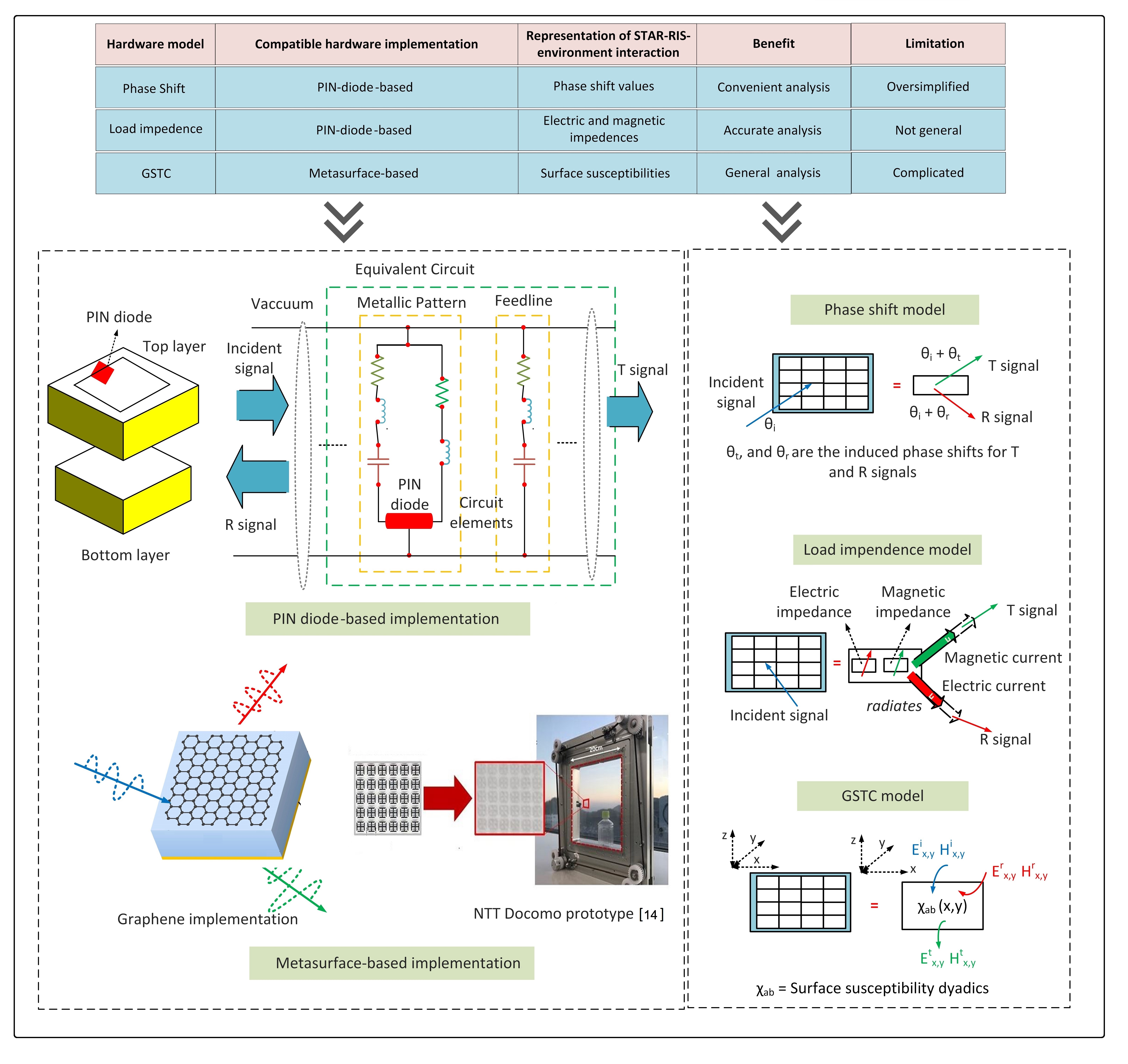}
\caption{Conceptual comparison of hardware implementations and modeling for STAR-RISs.
}
\label{fig2a}
\end{figure*}

\subsection{Hardware Implementations and Modeling}

{The conceptual comparison and compatibility of physical implementations and hardware models of the STAR-RISs are illustrated in Fig. \ref{fig2a}.}

\subsubsection{{PIN Diode- and Metasurface-based Implementations}}

From a manufacturing perspective, {the full-space manipulation of T and R signals demands different hardware implementations for STAR-RIS}. Depending on the tuning mechanism, hardware implementations can be divided into positive intrinsic negative (PIN) diode- and metasurface-based implementations. {PIN diode-based implementation relies on the PIN diodes distribution over two symmetrical layers. More precisely, the amplitude and phase variations of the T and R signals can be adjusted by the ON/OFF states of PIN diodes, which are controlled via DC voltage signals (biasing) \cite{ref14}.} Some other detail is as follow:
\begin{itemize}
\item \textit{Operating frequency}: {Generally used for} lower frequencies (10 kHz up to 1 GHz).
\item \textit{Advantage}: {Most common hardware realization} owing to its reduced cost, size, weight, and power consumption, and its voltage-controlled operating mechanism.
\item \textit{Disadvantage}: {Challenge to achieve} the independent T and R signals with a limited number of PIN diodes.
\end{itemize}

Metasurface-based implementation {uses a material whose EM properties (e.g., permittivity or conductivity) can be modified for effective electric and magnetic responses.} In this context, the prototype dynamic metasurface designed by NTT Docomo Inc. can dynamically control 28-GHz 5G radio signals. A prototype metasurface can successfully manipulate T and R signals in a highly transparent and compact package; therefore, it is suitable for unobtrusive use in buildings and vehicles. In addition, graphene has excellent EM properties in terms of carrier mobility and electrical conductivity, thereby enabling different potential applications in photonic devices. The conductivity of graphene can be dynamically modulated (electrically or magnetically). Therefore, tunable R and T coefficients can be achieved. {Graphene-based STAR-RISs} have excellent tunable response and low fabrication cost.

\subsubsection{{Modeling for Hardware Implementations}}

STAR-RISs differ from conventional RISs in terms of hardware design, physics principles, and communication system design. The development of a unified model for analyzing the reconfigurability of elements in different hardware implementations is still in its infancy. The physics-compliant hardware modeling to reflect the arbitrary EM field interaction with a STAR-RIS and characterize the T and R coefficients per element is an intrinsically complex problem; however, it is required to facilitate research on 6G STAR-RISs. In this regard, we present phase shift, load impedance, and generalized sheet transition conditions (GSTC) models to discuss the necessity and prerequisites of hardware characteristics and study the boundary conditions of the EM field at STAR-RISs. The phase shift model characterizes the response of the EM wave in terms of phase-shift values. In the load impedance model, the radiated T and R fields are characterized by electric and magnetic impedances. The GSTC model provides an analytical (algebraic) formulation of the electric and magnetic polarization densities as a function of macroscopic fields and dyadic surface susceptibilities. The aforementioned hardware models are based on different assumptions and approximations; however, they can enable a suitable insightful analysis.

\subsubsection{Open Issues and Future Directions}
{There is an inherent tradeoff between independent controls and design complexity for STAR-RISs. Therefore, novel hardware implementation and manufacturing solutions are required to make STAR-RISs more scalable and affordable while maintaining their tuning capabilities and independent phase and amplitude controls.}

\subsection{Practical Hardware Limitations for STAR-RISs}

Because of the complex nature of the interaction between STAR-RISs and the environment, ideal hardware realizations and oversimplified hardware models have been initially adopted in research studies. This limits the design and performance optimization in practice. The development of 6G STAR-RISs while addressing practical hardware limitations will be a challenge; therefore, further research is required. The practical hardware constraints, including the maximum number of elements, coupled phase shifts, phase-dependent amplitude, discrete (quantized) amplitude and phase shifts, and correlated channels, need to be considered.

\subsubsection{Coupled Phase Shifts}

The amplitude and phase shift variations for T and R users are generally chosen as constrained (law of energy conservation) and independent, respectively. However, independent T and R coefficients are non-trivial to realize for purely passive lossless STAR-RISs. We next list some details about coupled phase shifts:
{\begin{itemize}
\item In the ES protocol, the coupled phase control is required in general. For semi-passive or active STAR-RISs, however,  the independent phase shift is possible at the expense of increased manufacturing cost and reduced scalability.
\item In the TS and MS protocols, only the independent phase control is required due to separate T and R modes. However, such operations entail a higher hardware complexity due to frequently updating T and R coefficients.
\end{itemize}}

\subsubsection{Phase-Dependent Amplitude}

The amplitude response for conventional RISs is generally assumed to be independent of the phase-shift response. However, such ideal control is only possible for circuit implementations with zero energy dissipation. Similar to conventional RISs, the amplitude adjustments for T and R users introduced by a STAR-RIS will also be described in terms of non-uniform functions of the respective phase shifts. Specifically, each amplitude exhibits a minimum value for the phase shift near zero and a maximum value for the phase shift around $-\pi$ or $\pi$. It is important to characterize the intertwined relationship between the amplitude and phase-shift responses and capture the phase-dependent amplitude variations in element-wise T and R coefficients.

\subsubsection{Discrete Amplitude and Phase Shift}

The discrete phase shift and amplitude selection limits the ability of STAR-RISs to adjust the wireless environment and degrades the system performance. However, the continuous responses provide an upper-bound (and simpler) analysis at the expense of a high implementation cost owing to the high-resolution phase shift and amplitude controllers. The selection of the minimum control bits for quantized phase shift and amplitude levels per element to provide a cost-effective yet nearly-continuous shifting-based performance is vital.

\subsubsection{Correlated STAR-RIS Channels}
{The spatial correlation cannot be avoided in an isotropic scattering environment. Although independent and identically distributed (i.i.d.) Rayleigh model is not appropriate for sub-wavelength-sized elements, it can be used in design methodologies to provide an upper-bound, simpler, and tractable performance analysis.}

\subsubsection{Open Issues and Future Directions}
{The aforementioned hardware limitations complicate design and analysis. The development of 6G STAR-RISs can be summarized as follows:
\begin{itemize}
\item Assuming coupled phase shifts, phase-dependent amplitudes, and discrete amplitudes and phase shifts for T and R users, the investigation of  the STAR-RIS designs as well as the characterization of performance degradation will be challenging.
\item Spatially correlated Rayleigh fading provides a valid model and accurate insights under isotropic scattering while deriving performance metrics. The research on STAR-RISs can be first established using the simpler i.i.d. model for later extension to spatially correlated channels.
\end{itemize}}

\begin{table*}[t!]
{\caption{ Details of the optimization methods for STAR-RISs.}
\centering
\begin{tabular}{| p{3cm} |p{1.4cm}| p{3.1cm} | p{3.5cm} |   p{5.4cm} |}
\toprule

\textbf{Problem description}	& \textbf{{Divisions}} &   \textbf{Optimization methods}&   \textbf{Benefits}&   \textbf{Limitations} \\
\midrule

--  Non-convex (e.g., mixed-integer or NP-hard) problems {due to} the coupled variables (e.g., precoding and T and R coefficient matrices) and theoretic fundamental constraints (e.g., energy conservation and coupled phase-shift constraints in the ES protocol, and unit modulus constraints in the TS and MS protocols)&

  \;\;\;\;\;\;\;\
  
   \;\;\;\;\;\;\;\
   
    \;\;\;\;\;\;\;\
    
     \;\;\;\;\;\;\;\
     
      \;\;\;\;\;\;\;\
      
       \;\;\;\;\;\;\;\

Analytical approach

&

 \;\;\;\;\;\;\;\

  \;\;\;\;\;\;\;\

   \;\;\;\;\;\;\;\

    \;\;\;\;\;\;\;\
    
     \;\;\;\;\;\;\;\

 Iterative/search/alternating-based methods (e.g., AO, BCD, SCA, and SDR)

 &
  \;\;\;\;\;\;\;\

   \;\;\;\;\;\;\;\
   
    \;\;\;\;\;\;\;\
    
     \;\;\;\;\;\;\;\

  \;\;\;\;\;\;\;\

-- Acceptable sub-optimal or approximate solutions

-- Flexible performance-complexity trade-off

&

-- {Rely on} the mathematical models

-- Long execution time and high computational complexity for large-scale systems; not suitable for resource-limited and delay-sensitive applications

-- {Perfectly-known environment (e.g., availability of CSI) is required}

-- Unstable sometimes as the performance depends on the designed strategy

-- Not applicable to the practical scenarios {due to} simple analysis assumptions, e.g., static users

-- {Challenge to cope with a dynamic and heterogeneous environment} \\\cline{2-5}

 \;\;\;\;\;\;\;\

 \;\;\;\;\;\;\;\

   \;\;\;\;\;\;\;\
  
    \;\;\;\;\;\;\;\
   
-- 6G networks also generate high-dimensional /nonlinear/multi-objective problems &

  \;\;\;\;\;\;\;\

   \;\;\;\;\;\;\;\
   
    \;\;\;\;\;\;\;\
    
     \;\;\;\;\;\;\;\
     
      \;\;\;\;\;\;\;\
    
   \;\;\;\;\;\;\;\

  ML approach&

  \;\;\;\;\;\;\;\

-- Supervised learning (e.g., regression and classification)

-- Unsupervised learning (e.g., clustering and association analysis)

-- Federated learning

-- {DL (e.g., deep neural network)}

-- {DRL (e.g., Q-learning,
DDPG and policy gradient) }&

-- Do not completely rely on the mathematical models

-- Can manage the lack of the environment statistical information

-- Applicable under stringent computational time constraints due to high-speed graphical processing units

-- Suitable for the large-scale networks as well as dynamic and uncertain environment &
  
 -- Challenge to design an approach with
appropriate training data set in terms of availability, quality, and sufficiency

-- Challenge to design an approach with fast
convergence rate (learning speed)

-- Often presents high computational complexity

-- Selection of an approach applicable to large state and action spaces

-- Selection of an approach suitable for joint discrete and continuous state spaces (as both discrete and continuous parameters exist)

-- Selection of a relatively stable approach (i.e., insensitive to parameter setting)  \\
\bottomrule
\end{tabular}
\label{table4}}
\end{table*}

\subsection{Optimization Designs for STAR-RISs}

{The T and R coefficients design under theoretical fundamental constraints of STAR-RISs constitutes a different mathematical problem to conventional RISs having a single coefficient}  under a unit modulus constraint. The intractable objective function with complex constraints turns an optimization problem of STAR-RIS into a non-trivial task. {More precisely, the problems become non-convex because of the coupled variables, e.g., precoding and STAR-RIS coefficient matrices, and the theoretical fundamental constraints per element.} 
 Furthermore, complex topologies and non-linear components in 6G also give rise to high-dimensional problems. Therefore, the optimization designs of conventional RISs need to be extended, and effective solving methods are required for 6G STAR-RISs. Exact methods (e.g., linear programming, dynamic programming, and exhaustive search) can provide an optimal solution in a limited time for simpler problems only. {However, 6G STAR-RIS optimization problems are diverse and complex.} The solution and execution time via exact methods are unacceptable for large problems. As shown in Table \ref{table4}, {the top-level taxonomy of optimization theory (analytical) and ML covers the optimization techniques for 6G STAR-RISs.}

Generally, analytical approaches adopt iterative-based methods. Specifically, originally complicated problems are approximated into simpler sub-problems with convex relaxations to find a sub-optimal solution. The transformed sub-problems become tractable with decoupled optimization variables and can be solved with the guarantee of convergence to a stationary point of the original problem. In this context, iterative algorithms and alternative optimization techniques, including the penalty method, successive convex approximation (SCA), semidefinite relaxation (SDR), alternative optimization (AO), BCD, can be exploited. The second optimization approach implies the use of ML, which transforms wireless communications through its learning capability and large search space. By exploring the training data, ML-based approaches can directly learn solutions without explicit channel models while efficiently solving complex optimization problems of 6G STAR-RIS systems. {For example, the coupled T and R signals for an ES protocol requires a high-dimensional continuous and discrete control policy which can be developed using deep reinforcement learning (DRL) algorithms \cite{ref15}. Furthermore, DRL algorithms can also handle the multi (conflicting) objective problems, e.g., coverage and capacity optimization.}

\subsubsection{Open Issues and Future Directions}
{The future optimization design aspects of 6G STAR-RISs are as follows:
\begin{itemize}
\item Performance tradeoff on both sides of a STAR-RIS is an open research issue. The beamforming and deployment (location) optimization problems with users in the T and R half-spaced regions are more challenging, especially for realistic multi-user scenarios. 
\item The challenge for analytical approaches is to manage a highly dynamic wireless propagation environment, whereas for ML approaches it is to introduce a low-complex, accurate, and stable learning algorithm. With certain advantages and disadvantages, the selection of an appropriate approach depends on the design scenarios and communication objectives.
\end{itemize}}

\section{Conclusions}
\label{section4}
{This article aims to facilitate STAR-RIS implementations in 6G era. We study STAR-RIS designs for the simultaneous or consecutive T and R modes with coupled or independent controls. Specifically, the CSI acquisition, i.e., channel estimation, methods are presented. The comparison and compatibility of the physical implementations and hardware modeling, and the near- and far-field channel modeling are described. Finally, the high-dimensional optimization problems due to the STAR-RIS constraints and complex 6G networks are discussed. The future directions include the validation of proposed design solutions and insights with the hardware platforms and experimental measurements, and their extension in complex situations, e.g., mmWave/THz communications, hardware impairments, multilayer/multicell networks, and randomness of multiple users.}

\section*{Acknowledgement}
This research was supported by the Basic Science Research Program through the National Research Foundation of Korea (NRF) funded by the Ministry of Education(MOE) (NRF-2022R1I1A1A01071807, 2021R1I1A3041887), by the Institute of Information \& communications Technology Planning \& Evaluation (IITP) grant funded by the Korea government (Ministry of Science \& ICT (MSIT)) (2022-0-00704, Development of 3D-NET Core Technology for High-Mobility Vehicular Service), and by Korea University Grant.

\newpage
\section{Biographies}
\begin{IEEEbiographynophoto}{Waqas Khalid} received the Ph.D. degree in information and communication engineering from Yeungnam University, Gyeongsan, South Korea, in 2019. He is currently a Research Professor with the Institute of Industrial Technology, Korea University, Sejong, South Korea.
\end{IEEEbiographynophoto}

\begin{IEEEbiographynophoto}{Zeeshan Kaleem} (Senior Member, IEEE) received the Ph.D. degree in electronics engineering from Inha University, South Korea, in 2016. He is currently an Assistant Professor with the Electrical and Computer Engineering Department, COMSATS University Islamabad, Wah Campus. 
\end{IEEEbiographynophoto}

\begin{IEEEbiographynophoto}{Rehmat Ullah}  received the Ph.D. degree in electronics and computer engineering from Hongik University, South Korea, in 2020. He is currently a Lecturer (Assistant Professor) at the Cardiff School of Technologies, Cardiff Metropolitan University, UK. 
\end{IEEEbiographynophoto}

\begin{IEEEbiographynophoto}{Trinh Van Chien} (Member, IEEE) received the Ph.D. degree in communication systems from Link\"oping University, Sweden, in 2020. He is now a lecturer at the School of Information and Communication Technology, Hanoi University of Science and Technology, Vietnam. 
\end{IEEEbiographynophoto}

\begin{IEEEbiographynophoto}{Song Noh} (Member, IEEE) received the Ph.D. degree in electrical and computer engineering from Purdue University, West Lafayette, IN, USA, in 2015. Since September 2018, he has been with the Department of Information and Telecommunication Engineering, Incheon National University, Incheon, Korea, where he is now an Associate Professor.
\end{IEEEbiographynophoto}

\begin{IEEEbiographynophoto}{Heejung Yu} (Senior Member, IEEE) received the Ph.D. degree in electrical engineering from the Korea Advanced Institute of Science and Technology (KAIST), Daejeon, South Korea, in 2011. He is currently a Professor with the Department of Electronics and Information Engineering, Korea University, Sejong, South Korea. 
\end{IEEEbiographynophoto}

\begin{thebibliography}{1}
\bibliographystyle{IEEEtran}



\bibitem{ref1} W. Khalid {\it{et. al.}}, ``RIS-Aided Physical Layer Security with Full-Duplex Jamming in Underlay D2D Networks," {\it{IEEE Access}}, vol. 9, Jul. 2021, pp. 99667--99679.



\bibitem{ref2} X. Mu {\it{et. al.}}, ``Simultaneously Transmitting and Reflecting (STAR) RIS Aided Wireless Communications," {\it{IEEE Trans. Wireless Commun.}}, vol. 21, no. 5, May 2022, pp. 3083--3098.


\bibitem{ref3} H. Niu {\it{et al.}}, ``Weighted Sum Rate Optimization for STAR-RIS-Assisted MIMO System," {\it{IEEE Trans. Veh. Technol.}}, vol. 71, no. 2, Feb. 2022, pp. 2122--2127.



\bibitem{ref4} A. Papazafeiropoulos {\it{et al.}}, ``Coverage Probability of STAR-RIS Assisted Massive MIMO Systems with Correlation and Phase Errors," {\it{IEEE Wireless Commun. Lett.}}, vol. 11, no. 8, Aug. 2022, pp. 1738--1742.


\bibitem{ref5} C. Wu {\it{et al.}}, ``Channel Estimation for STAR-RIS-Aided Wireless Communication," {\it{IEEE Commun. Lett.}}, vol. 26, no. 3, Mar. 2022, pp. 652--656.


\bibitem{ref6} W. Khalid {\it{et. al.}}, ``Rate-Energy Tradeoff Analysis in RIS-SWIPT Systems with Hardware Impairments and Phase-Based Amplitude Response," {\it{IEEE Access}}, vol. 10, Mar. 2022, pp. 31821--31835.


\bibitem{ref7} H. Lu  {\it{et al.}}, ``Aerial Intelligent Reflecting Surface: Joint Placement and Passive Beamforming Design With 3D Beam Flattening," {\it{IEEE Trans. Wireless Commun.}}, vol. 20, no. 7, Jul. 2021, pp. 4128--4143.



\bibitem{ref8} T. Hou {\it{et al.}}, ``A Joint Design for STAR-RIS Enhanced NOMA-CoMP Networks: A Simultaneous-Signal-Enhancement-And-Cancellation-Based (SSECB) Design," {\it{IEEE Trans. Veh. Technol.}}, vol. 71, no. 1, Jan. 2022, pp. 1043--1048.

\bibitem{ref9} Y. Guo {\it{et al.}}, ``Energy-Efficient Design for a NOMA Assisted STAR-RIS Network with Deep Reinforcement Learning," Nov. 2021. [Online]. Available at: https://arxiv.org/abs/2111.15464.

\bibitem{ref10} J. Xu {\it{et al.}}, ``STAR-RISs: Simultaneous Transmitting and Reflecting Reconfigurable Intelligent Surfaces," {\it{IEEE Commun. Lett.}}, vol. 25, no. 9, Sep. 2021, pp. 3134--3138.





\bibitem{ref11} G. M. Minopoulos {\it{et al.}}, ``Exploitation of Emerging Technologies and Advanced Networks for a Smart Healthcare System," {\it{Appl. Sci.}}, vol. 12, no. 12, Jun. 2022, p. 5859.


\bibitem{ref12} S. Noh {\it{et al.}}, ``Channel Estimation Techniques for RIS-Assisted Communication: Millimeter-Wave and Sub-THz Systems," {\it{IEEE Veh. Technol. Mag.}}, vol. 17, no. 2, Jun. 2022,  pp. 64--73.



\bibitem{ref13} L. Wei {\it{et al.}}, ``Channel Estimation for RIS-Empowered Multi-User MISO Wireless Communications," {\it{IEEE Trans. Commun.}}, vol. 69, no. 6, Jun. 2021, pp. 4144--4157.



\bibitem{ref14} H. Zhang {\it{et al.}}, ``Intelligent Omni-Surfaces for Full-Dimensional Wireless Communications: Principles, Technology, and Implementation," {\it{IEEE Commun. Mag.}}, vol. 60, no. 2, Feb. 2022, pp. 39--45.




\bibitem{ref15} R. Zhong {\it{et al.}}, ``Hybrid Reinforcement Learning for STAR-RISs: A Coupled Phase-Shift Model Based Beamformer," {\it{IEEE J. Sel. Areas Commun.}}, vol. 40, no. 9, Sep. 2022, pp. 2556--2569.


\end{thebibliography}
\end{document}